\documentclass[epj]{svjour}
\usepackage{graphicx}
\usepackage{xspace}
\usepackage{amsmath,amssymb}
\usepackage[T1]{fontenc}
\usepackage[latin9]{inputenc}
\usepackage{amstext}
\usepackage{esint}
\usepackage[english]{babel}

\def\kbar{{\mathchar'26\mkern-9mu k}}

\begin{document}

\title{Kicked-rotor quantum resonances in position space: Application to
situations of experimental interest}

\author{Maxence Lepers
     \thanks{Present address: Laboratoire Aim{\'e} Cotton, Universit{\'e} Paris-Sud, Bat. 505, Campus d'Orsay, F-91405 Orsay Cedex, France}
   \and V{\'e}ronique Zehnl{\'e}
   \and Jean-Claude Garreau
}

\institute{Laboratoire de Physique des Lasers, Atomes et Mol{\'e}cules, Universit{\'e} Lille 1 Sciences et Technologies, CNRS; F-59655 Villeneuve d'Ascq Cedex, France}

\authorrunning{M. Lepers, V. Zehnl{\'e}, J.-C. Garreau}
\titlerunning{Kicked-rotor quantum resonances in position space...}

\date{\today}

\abstract{
In this work we apply the formalism developed in {[}M. Lepers \emph{et
al}., Phys. Rev. A \textbf{77}, 043628 (2008)] to different initial
conditions corresponding to systems usually met in real-life experiments, and calculate the observable quantities that can be used
to characterize the dynamics of the system. The position space point
of view allows highly intuitive pictures of the physics at play.
}

\maketitle

\section{Introduction}

Over the last 30 years, the \emph{kicked rotor} has played a central
role in the field of \emph{quantum chaos}, which is defined as the
study of quantum systems whose classical counterpart is chaotic \cite{StoeckmannChaos,GutzwillerChaos,Lichtenberg:ChaoticDynamics:82}.
The first atom-optics experimental realization of its quantum version
in 1995 \cite{Raizen:QKRFirst:PRL95} was obtained by submitting laser-cooled
atoms to a periodic series of {}``kicks'' of a laser standing wave
which is put on periodically for a very short time compared to the
atom dynamics. It has been shown that, under quite general conditions,
this system displays a characteristic behavior named {}``dynamical
localization'', that is, an asymptotic saturation of its average
kinetic energy, that can be attributed to the existence of destructive
interferences in the system. An impressive number of experimental
results followed this pioneering work \cite{Amman:LDynNoise:PRL98,Raizen:KRClassRes:PRL98,DArcy:AccModes:PRL99,AP:Bicolor:PRL00,Monteiro:DoubleKick:PRL04,Renzoni:RatchetOptLat:PRL04,AP:Reversibility:PRL05,AP:PetitPic:PRL06,AP:Anderson:PRL08},
spanning domains like quantum transport \cite{2008-dana-prl:QRRatchet-Exp},
measurements of the gravitational constant \cite{DArcy:GravityQuantRes:PRL04},
condensed matter and quantum phase transitions with, recently,
the first experimental observation of the metal-insulator Anderson
transition with matter waves \cite{AP:Anderson:PRL08,AP:AndersonLong:PRA09,AP:AndersonCritic:PRL10}.

Dynamical localization is not the only interesting phenomenon
displayed by the kicked rotor. Another striking feature of the quantum
kicked rotor (QKR) is the existence of \emph{quantum resonances} (QRs)%
\footnote{Not to be confused with classical resonances or accelerator modes.
These are observed for particular values of the stochasticity parameter
(see below) $K\approx2\pi n$ and of the initial conditions.%
}, which arise if the kicking period matches the evolution of
the quantum phase of the atoms. A QR is characterized by a \emph{ballistic}
behavior in which the velocity increases linearly and the average
kinetic energy quadratically with time, in sharp contrast with dynamical
localization. Since the pioneering theoretical works \cite{Izrailev:LocDyn:PREP90,Shepel:QuantResPosSpace:TMP80},
the QRs have been the subject of an impressive amount of work, both
theoretical
\cite{fishman1993,2000-sokolov-prl:QR:SUq,2000-sokolov-pre:QuasiHamilt,Wimberger:QuantRes:NL03,Dana:GeneralQR:PRE06,1995-dana-prl:QR:LocDyn,2004-wimberger-prl:QR:Scal,AP:QuantumRes:PRA08,2007-saunders-pra:QR:FinTemp,2009-guarneri-ahp:QR:Spectrum}
and experimental
\cite{Raizen:QKRFirst:PRL95,raizen1999,Darcy:QRes:PRL01}
\cite{2000-oskay-oc:QR:BallistPeaks,2005-sadgrove-prl,2004-sadgrove-pre,2005-wimberger-pra,Phillips:HighOrderQuantResBEC:PRL06,Steinberg:HighOrderQRes:PRL07,2008-sadgrove-pre:ScalLaw},
including the effect of gravity (the so-called quantum accelerator
modes) \cite{DArcy:AccModes:PRL99,Fishman:KRQuantumRes:PRL02,DArcy:HighOrderQRes:PRL03,Summy:QKAPhaseSpace:PRL06,2003-fishman-jsp:QR:GravTh,DArcy:QuantumStability:PRL03,2008-halkyard-pra:PowLaw},
directed motion (or quantum ratchets) \cite{2005-lundh-prl-QRRatchets,2007-sadgrove-prl:QRRatchet,2009-sadgrove-prl:QRRatch-PhaNoise,Sadgrove:TrasportQR:NJP09,2008-dana-prl:QRRatchet-Exp},
or the effects of interatomic interactions in a Bose-Einstein condensate
(BEC) \cite{Monteiro:QRBEC:PRL09,Raizen:QRBEC:PRL06}.

The main goal of activity around QRs is to calculate and measure the average
momentum and kinetic energy, and possibly control the dynamics,
\textit{e.g.} by constructing \emph{ad hoc} position or momentum distributions.
The present article is placed in this mainstream. We apply analytical
methods previously developed in \cite{AP:QuantumRes:PRA08}, validated
by comparison with numerical simulations, to the mean momentum and
kinetic energy of the QKR in situations often encountered in the laboratory.
Some of these results have already been presented in other works,
but our approach and the intuitive picture of the dynamics
sheds a new light on these complex behaviors.

Systems that are periodic in space, as the kicked rotor is, conserve
quasimomentum. This means that its eigenstates are combs displaying
the same periodicity as the potential, both in the position and in
the momentum space, the Fourier transform of a comb being a comb.
A comb-shaped spatial wave function, in the absence of a potential,
has the striking property of \emph{resuming
its initial (strongly localized) shape} after some characteristic
time, an effect known in optics as the Talbot effect \cite{Berry:TalbotEffect:JMO99}.
The QRs of the kicked rotor are a quantum version of the Talbot effect.
In \cite{AP:QuantumRes:PRA08} we performed an analysis of such effect
in the position space, which allowed us to simply relate the QR dynamics
to \emph{local} properties of the potential at the positions where
the wave function reconstructs. The QRs' ballistic behavior then arises
as a consequence of the fact that the wave function focuses, before
each kick, at the same position with respect to the potential. The
system thus receives the same amount of momentum from each
kick, and the effect of successive kicks adds constructively. The resulting dynamics,
although based on intereferences, can be interpreted in terms of purely
mechanical arguments.

However, the initial conditions of real experiments are seldom comb-shaped
functions. The initial state of a Bose-Einstein condensate (BEC),
for example, can be modeled as a sharp superposition of plane waves
around zero momentum. The hotter atom cloud from a magneto-optical trap
is represented as an incoherent mixture of plane waves. The purpose
of the present article is to show that the picture presented in \cite{AP:QuantumRes:PRA08}
is also valid in those situations. The main point is that an arbitrary
initial wave function can be viewed as a superposition of comb-shaped
states, which, \emph{in the specific case of QRs}, evolve independently
from each other. We consider in the present work only the case of
the so-called \emph{simple} quantum resonances, in which the reconstructed
wave packet is identical to the initial one. The more complex case
of the so-called ``higher-order'' quantum resonances, in which
the initial wave packet reconstructs into many replicas, will be discussed
in a future work. Thus in what follows, unless otherwise stated, it
is understood that the term \emph{quantum resonance} means \emph{simple
quantum resonance}.

The article is organized as follows. In section \ref{sec:rappel},
we recall the essential features of our approach to QRs in position
space. We then generalize these results to initial conditions of experimental
interest: After considering the case of a plnae wave (sec. \ref{sec:Plane-wave}),
which is the basic component to describe experimental situations,
we address the case of a coherent superposition of two plane waves
(sec. \ref{sub:2PW}), which is used to interpret the recent experiments
on quantum transport and ratchets, the case of a narrow momentum distribution,
which is characteristic of a BEC (sec. \ref{sub:Narrow}); finally,
for the sake of completeness, we briefly discuss the {}``trivial''
case of a broad initial momentum distribution, characteristic of a
thermal cloud (sec. \ref{sub:Broad}). In all these cases, our picture
of the evolution seen in position space allows a simple interpretation
of the dynamics. Section \ref{sec:Conclusion} draws the conclusions
of this work.

\section{Quantum resonances in position space} \label{sec:rappel}

The atomic kicked rotor is obtained by placing laser-cooled atoms
of mass $M$ and momentum $p$ in a standing wave formed by two counterpropagating
(along the $x$-axis) laser beams of wave number $k_{L}=2\pi/\lambda_{L}$.
The atoms are thus submitted to a mechanical potential called ``optical
potential'' $V(x)=V_{0}\cos\left(2k_{L}x\right)$ formed by the standing
wave. If the standing wave is modulated in the form of periodic pulses
(period $T$) of duration $\tau$ very short at the timescale of the
atom dynamics%
\footnote{That is $p\tau/M\ll\lambda_{L}$ for all relevant values of the atomic
momentum $p$.}, one obtains the well-known kicked rotor Hamiltonian
\begin{equation}
H=\frac{P^{2}}{2}+K\cos X\sum_{n}\delta(t-n),
\end{equation}
where we used reduced variables \cite{Raizen:QKRFirst:PRL95} in which
time is measured in units of the kicking period $T$, $X=2k_{L}x$, $P=\kbar p/2\hbar k_{L}$,%
\footnote{As the atoms move on a straight line, $X\in]-\infty;+\infty[$ and $P\in]-\infty;+\infty[$.%
}
$K=\kbar V_{0}\tau/\hbar$, where $\kbar=4\hbar k_{L}^{2}T/M$ plays
the role of a reduced Planck constant %
\footnote{In the sense that the corresponding Schr{\"o}dinger equation is $i\kbar\partial\psi/\partial t=H\psi$.%
}. The parameter $\kbar$, which plays a critical role in what follows,
can be experimentally tuned by changing the kicking period $T$. 

The quasimomentum $\beta$ is defined by the relation
\begin{equation}
P=\left(n+\beta\right)\kbar,
\end{equation}
where $n$ is an integer and $\beta\in[-\frac{1}{2};\frac{1}{2})$
is in the first Brillouin zone. The spatial periodicity of the potential
implies that $\beta$ is a constant of motion%
\footnote{The conservation of quasimomentum is a general property of periodic
potentials, but the case of the atomic kicked rotor allows a completely
different interpretation. Interacting with a far-detuned standing
wave, an atom can only absorb and re-emit photons by stimulated emission.
Four kinds of process are possible: Two of them correspond to the
absorption and emission of a photon in the same propagative wave forming
the standing wave; in this case the overall momentum exchanged between
the atom and field is zero. Two other process correspond to the absorption
of a photon in one of the propagative waves followed by the emission
of a photon in the \emph{other} propagative wave, in which case the
momentum exchanged is two times the photon momentum, that is $2\hbar k_{L}$.
Starting from a well-defined momentum $p_{0}$ only momentum states
of the form $2n\hbar k_{L}$ (or $n\kbar$ in our units) are populated.
The conservation of quasimomentum can thus be deduced from this {}``microscopic''
argument, not directly related to the periodicity of the potential.%
}. In position space, wave functions $\psi_{\beta}(X)$ corresponding
to a well-defined quasimomentum have the well-known Bloch-wave structure
\begin{equation}
\psi_{\beta}(X)=e^{i\beta X}u_{\beta}(X),
\end{equation}
where $u_{\beta}(X)$ has the same spatial period $2\pi$ as the potential
\footnote{The atom wave function is thus $\psi(X)=\int_{-1/2}^{1/2}d\beta\psi_{\beta}(X)$%
}.

The dynamical evolution of the atoms is governed by the one-(temporal)
period evolution operator of the quantum kicked rotor
\begin{equation}
U=\exp\left(-i\frac{K}{\kbar}\cos X\right)\exp\left(-i\frac{P^{2}}{2\kbar}\right),
\label{eq:EvolOper}
\end{equation}
where the first term on the right-hand side corresponds to the effect
of an instantaneous kick and the second to a free propagation between
two kicks. The factorization is allowed because the kicks are short
compared to the atom dynamics, so that one can neglect the contribution
of the free-evolution term during this short time. (General) quantum
resonances appear if $\kbar$ is a rational multiple of $4\pi$, \emph{i.e.}
$\kbar=4\pi r/s$. In physical units, this corresponds to a kicking
period $T=rT_{T}/s$, where $T_{T}=\pi M/\hbar k_{L}^{2}$ is a characteristic time, called
the Talbot time. Given a state $\psi_{\beta}(X,t)$
of well-defined quasimomentum $\beta$, one can use Eq. (\ref{eq:EvolOper})
to calculate the wave function $\psi_{\beta}(X,t+1)$ one period later.
The condition for simple quantum resonances is $\kbar=2\pi\ell$,
with $\ell$ integer, which leads to a particularly simple recursion
relation {[}see \cite{AP:QuantumRes:PRA08}, Eq. (10)] for the wave
function:
\begin{eqnarray}
\psi_{\beta}(X,t) & = & e^{-iK\cos X/\kbar} \exp\left(i\kbar\beta(\beta+1)/2\right) \nonumber\\
 & \times & \psi_{\beta}\left(X-\kbar\left(\beta+1/2\right),t-1\right).\label{eq:SQR-rec-psi}
\end{eqnarray}
The term $\exp\left(i\kbar\beta(\beta+1)/2\right)$ in Eq. (\ref{eq:SQR-rec-psi})
is a global phase that does not affect the dynamics. The free evolution
between kicks thus preserves the shape of $\psi_{\beta}$ but translates
it, over a whole kick period, of\begin{equation}
v_{\beta}=\kbar\left(\beta+\frac{1}{2}\right).\end{equation}

In \cite{AP:QuantumRes:PRA08} we have considered the evolution of
an atom initially prepared in a Bloch wave perfectly localized in
a spatial period $[-\pi,\pi)$ %
\footnote{Let us stress that here ``localized'' means localized in the ``spatial''
first Brillouin zone $[-\pi,\pi)$, the full wave function is a comb
of delta functions. In the momentum space the wave function is also
a Dirac comb; the Heisenberg's uncertainty relations are thus fully satisfied.%
} and centered at a position $\xi\in[-\pi,\pi)$, that is
\begin{equation}
\chi_{\beta}(X-\xi)=e^{i\beta X}\sum_{j=-\infty}^{+\infty}\delta\left(X-\xi-2\pi j\right).
\label{eq:RQS-WF0}
\end{equation}
In what follows, $\chi_{\beta}$, which will play a central role in
our interpretation of QRs, will be referred to as a comb-shaped Bloch
wave (CSBW). In \cite{AP:QuantumRes:PRA08} we have shown that the
mean position and momentum evolution of a CSBW obeys a simple map
\begin{eqnarray}
X_{\beta,\xi}(t) & = & X_{\beta,\xi}(t-1)+v_{\beta}=\xi+v_{\beta}t\label{eq:RQS-map-X}\\
P_{\beta,\xi}(t) & = & P_{\beta,\xi}(t-1)+K\sin(\xi+v_{\beta}t)\nonumber \\
 & = & P_{\beta,\xi}(0)+K\sum_{s=1}^{t}\sin(\xi+v_{\beta}s),
\label{eq:RQS-map-P}
\end{eqnarray}
and the evolution of the mean kinetic energy turns out to be given
by
\begin{equation}
E_{\beta,\xi}(t)-E_{\beta,\xi}(0)=\frac{1}{2}\left(P_{\beta,\xi}^{2}(t)-P_{\beta,\xi}^{2}(0)\right).
\label{eq:RQS-map-E}
\end{equation}
Unlike the $\epsilon$-classical map, used to describe quasi-resonant
dynamics \cite{Fishman:KRQuantumRes:PRL02,2003-fishman-jsp:QR:GravTh},
Eqs. (\ref{eq:RQS-map-X})--(\ref{eq:RQS-map-E}) do not describe
a classical motion, as the change in position $X_{\beta,\xi}(t)-X_{\beta,\xi}(t-1)$
does not depend on momentum at kick $t-1$.

The CSBW in Eq. (\ref{eq:RQS-WF0}) is characterized by its quasimomentum
$\beta$ and by its {}``center'' $\xi$ which is determined by the
weights of each momentum component $n\kbar$. In the case of a quantum
resonance -- and only in this case -- the evolution between two subsequent
kicks just translates the wave packet of $v_{\beta}$; therefore,
for fixed $\beta$, contributions due to different initial values
of $\xi$ are not mixed, and each CSBW $\chi_{\beta}(\xi)$ evolves
independently of the others. An arbitrary initial state can be expanded
in CSBWs $\chi_{\beta}(X-\xi)$ {[}see Eq. (\ref{eq:RQS-WF0})] 
\begin{equation}
\psi_{\beta}(X,t=0)=\int_{-\pi}^{\pi}d\xi\psi_{\beta}(\xi,t=0)\chi_{\beta}(X-\xi)\,,\end{equation}\label{eq:psibeta}
with 
\begin{equation}
\psi_{\beta}(\xi,t=0)=\frac{1}{\sqrt{2\pi}}\sum_{n}e^{i(n+\beta)\xi}\widetilde{\psi}_{\beta}(n),\label{eq:psibetan}
\end{equation}
$\widetilde{\psi}_{\beta}(n)$ being the amplitude of the component
$n$ in momentum space.

In the following sections, we apply the above formalism to cases of
experimental interest where the initial state is not a CSBW, but either
a coherent or an incoherent mixture of CSBWs. As the CSBWs evolve independently,
we have just to find the weights $\pi_{\beta,\xi}$ of each CSBW in
the initial state, which then allows us to
directly calculate the averages 
\begin{eqnarray}
\overline{P^{k}}(t)-\overline{P^{k}}(t=0) & = & \int_{-1/2}^{1/2}d\beta\int_{-\pi}^{\pi}d\xi\pi_{\beta,\xi} \nonumber\\
 & & \times \left(P_{\beta,\xi}^{k}(t)-P_{\beta,\xi}^{k}(t=0)\right)
\label{eq:Average}
\end{eqnarray}
with $k=1,2$, and $P^k_{\beta,\xi}$ given by Eqs. (\ref{eq:RQS-map-P})
and (\ref{eq:RQS-map-E}).  In the case
where the initial state is a coherent superposition of plane waves,
$\pi_{\beta,\xi}$ is given by
\begin{eqnarray}
\pi_{\beta,\xi} & = & \left|\psi_{\beta}(\xi,t=0)\right|^{2} \nonumber\\
 & = & \frac{1}{2\pi} \sum_{n,n'} e^{i(n-n')\xi} 
\widetilde{\psi}_{\beta}^{*}(n') \widetilde{\psi}_{\beta}(n) \,,
\label{eq:pi-psibeta}
\end{eqnarray}
while in the case of an incoherent mixture, $\pi_{\beta,\xi}$ is 
\begin{equation}
\pi_{\beta,\xi} = \frac{1}{2\pi} \sum_{n} 
\left|\widetilde{\psi}_{\beta}(n)\right|^{2} \,.
\label{eq:pi-rho}
\end{equation}

\section{Plane wave}\label{sec:Plane-wave}

Let us first apply the previous results to the simple case of a plane
wave. A plane wave of momentum $p_{\text{0}}$ has a well-defined
quasimomentum $\beta_{0}$ which satisfies $p_{0}=(n_{0}+\beta_{0})\kbar$,
but is completely delocalized in the real space, which means that
all values of $\xi$ are equiprobable on $[-\pi,\pi)$, hence $\pi_{\beta,\xi}=\delta(\beta-\beta_{0})/2\pi$.
From Eqs. (\ref{eq:RQS-map-P}) and (\ref{eq:Average}) one finds,
\begin{eqnarray}
\overline{P}(t) & = & \overline{P}(0) +
\frac{K}{2\pi} \int_{-\pi}^{\pi}d\xi \sum_{s=1}^{t} \sin\left(\xi+v_{\beta_0}s\right)\nonumber \\
 & = & p_{0}.\label{eq:RQS-P-op}\end{eqnarray}
The average kinetic energy is obtained in the same way:\begin{eqnarray}
\overline{E}(t) & = & \overline{E}(0) + \frac{K^{2}}{4\pi}\int_{-\pi}^{\pi}d\xi 
\left( \sum_{s=1}^{t} \sin(\xi+v_{\beta_0}s)\right)^{2}\nonumber \\
 & = & \frac{\kbar^{2}\left(n_{0}+\beta_{0}\right)^{2}}{2} 
 +\frac{K^{2}}{4}\frac{\sin^{2}(v_{\beta_0}t/2)}{\sin^{2}(v_{\beta_0}/2)},
\label{eq:RQS-E-op}
\end{eqnarray}
a result previously obtained in \cite{Wimberger:QuantRes:NL03}. If
$v_{\beta_0}=0[2\pi]$, where $x[2\pi]$ means {}``$x$ modulus $2\pi$'',
the dynamics is ballistic: $\overline{E}(t)=E(0)+K^{2}t^{2}/4$, whereas
if $v_{\beta_0}=\pi[2\pi]$ (e.g. $\kbar=2\pi,\beta_{0}=0$) the dynamics
is periodic (or {}``anti-resonant''), i.e. $\overline{E}(t)=\overline{E}(0)+K^{2}/4$
for $t$ odd and $\overline{E}(t)=\overline{E}(0)$ for $t$ even.

\begin{figure}
\begin{centering}
\includegraphics[width=8cm]{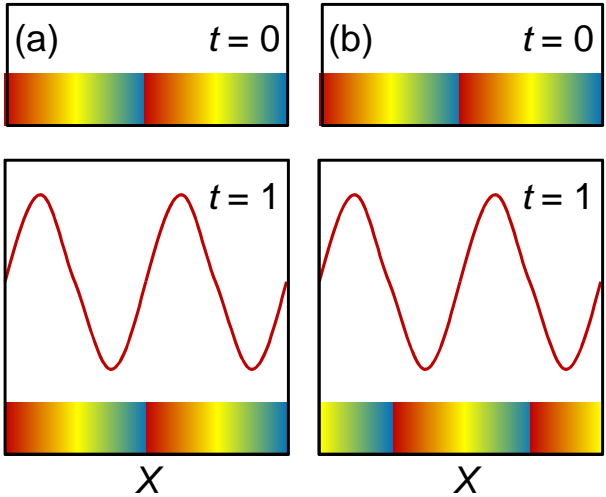}\caption{\label{fig:SRQ-E-op}(color on line) Dynamics of a zero-momentum plane
wave. The initial distribution, uniform in the position space (restricted
here to two potential wells), is divided into ``slices''  represented in different colors.
Panel (a) corresponds to the ballistic evolution of the quantum resonance
$\kbar=4\pi$. After a free evolution between kicks, each color has come
 back to the initial position (slices of same color are at the same position in
the bottom ``degrad{\'e}" as in the top one).
The subsequent kick is thus  applied to the \emph{same} spatial distribution, and each
slice receives the \emph{same} momentum from any kick. The
momentum transfers thus add constructively from kick to kick. Panel
(b) corresponds to the ``anti-resonance'' condition
$\kbar=2\pi$. In such case, each colored slice is shifted by half the lattice step by the
free evolution between kicks, and
thus receives from a given kick a momentum transfer of the
same magnitude, but of opposite direction with respect to the previous
kick, which leads to a periodic behavior.}

\par\end{centering}
\end{figure}

Figure \ref{fig:SRQ-E-op} gives a simple interpretation of this behavior
for a zero-momentum plane wave ($n_{0}=\beta_{0}=0$). On panels (a)
and (b), the initial position-space distribution is uniform, but we
have attributed a different color to each slice of it, so that one
can follow its evolution. Eq. (\ref{eq:RQS-map-X}) shows that the free
evolution between kicks results in a simple global shift of the wave function between
$t=0$ and $t=1$ kick. In the ballistic case (a), this shift is such
that all ``colors'' (or slices) come back to the same place at
$t=1$ (and also at $t=2,3,...$). Therefore, each part of the wave
function receives the same momentum transfer $K\sin \xi$ [for example, the yellow part receives
the maximum momentum transfer $+K$, see Eq.~(\ref{eq:RQS-map-P})]. As the wave function fills uniformly
in the potential wells when the kicks are applied, the average momentum
is always zero. But, the average kinetic energy grows quadratically,
since some ``colors'' are accelerated in the positive direction,
while others are accelerated in the negative direction. \emph{Although
the position-space distribution is uniform, the local effect of the
kicks remains visible on the average kinetic energy.} By contrast,
figure \ref{fig:SRQ-E-op} (b) illustrates the so-called anti-resonant
case, for which the position-space kicks, a given ``color''
receives a momentum equal to $-K\sin\xi$, whereas at even kicks, 
it receives $+K\sin\xi$, hence a periodic behavior.

\section{Superposition of two plane waves: inducing directed
motion} \label{sub:2PW}

It has recently been show that directed motion of the wave packet
(which is sometimes called a ``quantum ratchet'') can be
obtained by preparing a BEC in a coherent superposition of two momentum
states via a Bragg pulse \cite{2005-lundh-prl-QRRatchets,2009-sadgrove-prl:QRRatch-PhaNoise,Sadgrove:TrasportQR:NJP09}.
This yields a superposition of two plane waves: say, one centered at
$P=0$ and one centered at $P=-\kbar$, each plane wave being equally
populated. We present here an alternative, and hopefully more intuitive,
interpretation of this result using the position-space point of view.

A coherent superposition of two plane waves $(|0\rangle-ie^{i\phi}|-\kbar\rangle)/\sqrt{2}$
is characterized by a mean momentum equal to $-\kbar/2$ \cite{2007-sadgrove-prl:QRRatchet}.
The corresponding initial CSBW distribution $\pi_{\beta,\xi}$ is
obtained from
\begin{equation}
\widetilde{\psi}_{\beta}(n)=\frac{\delta(\beta)}{\sqrt{2}}\left(\delta_{n,0}-ie^{i\phi}\delta_{n,-1}\right)
\label{eq:RatchetPsi0}
\end{equation}
and Eq. (\ref{eq:psibetan}): 
\begin{equation}
\pi_{\beta,\xi}=\left|\psi_{\beta}(\xi,t=0)\right|^{2}=\frac{\delta(\beta)}{2\pi}\left(1-\sin\left(\xi-\phi\right)\right),
\label{eq:Ratchet-psi2}
\end{equation}
which is localized in quasimomentum and inhomogeneous in the position
representation. Performing the integral in Eq. (\ref{eq:Average})
we retrieve Eq. (5) of ref. \cite{2007-sadgrove-prl:QRRatchet}
\begin{equation}
\overline{P}(t)=-\frac{\kbar}{2}-\frac{Kt}{2}\cos\phi,
\label{eq:ratchet-P}
\end{equation}
which shows a directed motion, which is \emph{not} due to the non-zero
initial average momentum, but to the inhomogeneity of the position-space
distribution.
This point can be stressed by considering an initial superposition of two arbitrary  plane waves 
$\left(\left|n\kbar\right\rangle\right.$
$\left.-ie^{i\phi}\left|\left(n-1\right)\kbar\right\rangle \right)$
$/\sqrt{2}$. By applying Eqs. (\ref{eq:pi-psibeta}) and (\ref{eq:psibetan}), one obtains the same distribution $\pi_{\beta,\xi}$ as in Eq. (\ref{eq:Ratchet-psi2}).
For all value of $n$, the directed motion will be characterized by the same current, whatever the initial average momentum and kinetic energy. This remarkable property emphasizes the central role of $\pi_{\beta,\xi}$ on the dynamical evolution of the QKR.

In Eq. (\ref{eq:ratchet-P}), the direction of the motion is controlled by
the quantum phase. For $\phi=0$, the maximum of $\pi_{\beta,\xi}$
is located at $\xi=-\pi/2$, where the local momentum transferred
from the pulse reaches its minimal value $-K$ (cf. Fig. \ref{fig:SRQ-E-op}),
which creates a dominant current towards $-X$, thus corresponding
to a negative average momentum. The opposite situation $\phi=\pi$
corresponds to the case where the maximum of $\pi_{\beta,\xi}$ coincides
with the maximum positive transferred momentum ($+K$): the current
is maximum towards the $+X$ direction. The prefactor $1/2$ in the current
comes from the averaging over the
width of the spatial distribution, which means that some slices of
the distribution receive an amount of momentum which is smaller than
$K$ in absolute value. The amplitude of the current can thus be increased
by building a position-space distribution sharply localized on favorable
values of the spatial potential.

As pointed out in Ref. \cite{2005-lundh-prl-QRRatchets}, such kind
of quantum transport cannot be observed if the initial state is a plane wave. Our picture
of QRs makes this result easily understandable, as a plane wave equally
samples all parts of the kicking potential. This reasoning is straightforwardly
generalized to the arbitrary kicking potentials obtained by adding
higher harmonics of the spatial frequency $2k_{L}$ \cite{Monteiro:RatchetOptLatTh:PRL02}.

In the anti-resonant case $\kbar=2\pi\ell$ ($\ell$ odd), no transport
can be observed, even with an inhomogeneous distribution, as given
slice at position $\xi$ will always receive from kick $t$ a momentum
exactly opposite to the one received at kick $t-1$.

\section{Narrow initial momentum distribution} \label{sub:Narrow}

\begin{figure}
\begin{centering}
\includegraphics[width=8cm]{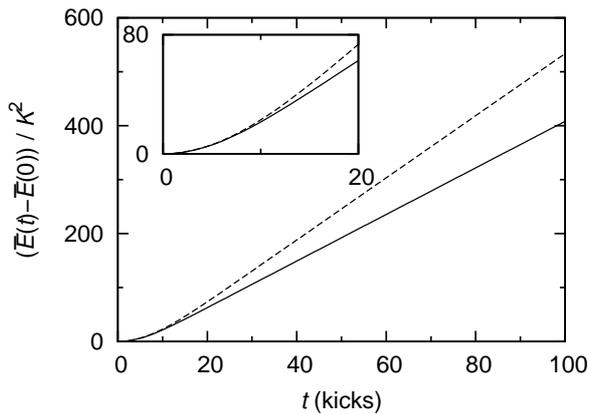}
\par\end{centering}

\caption{\label{fig:RQS-Emoy-t-balist}Time evolution of the average kinetic energy obtained
by integration of Eq. (\ref{eq:EvolOper}). The initial momentum distributions
are coherent Gaussian superpositions of plane waves, whose central quasimomenta $\beta=0$ are ballistic ($\kbar=4\pi$).
Their root mean squares, $\sigma=0.0115$ (solid lines) and $\sigma=0.00866$
(dashed lines), are such that $\Delta=0.04$ and 0.03 respectively (see text). 
Other parameter: $K=10$.}

\end{figure}

\begin{figure}
\begin{centering}
\includegraphics[width=8cm]{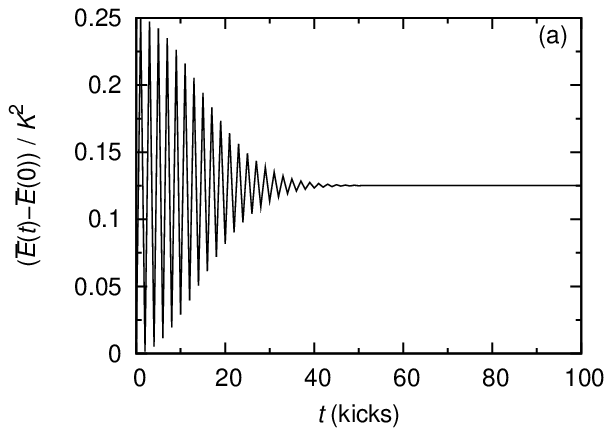}
\includegraphics[width=8cm]{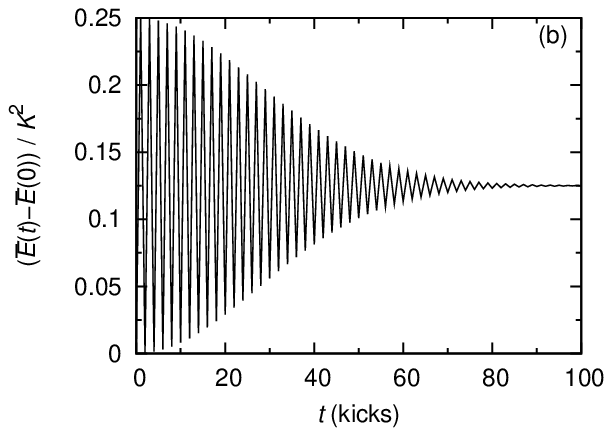}
\par\end{centering}

\caption{\label{fig:RQS-Emoy-t-antir}
Same as Fig.~\ref{fig:RQS-Emoy-t-balist}, except that the central component of
the initial momentum distributions is anti-resonant ($\kbar=2\pi,\beta=0$).
Their root mean squares are: (a) $\sigma=0.0115$ and (b) $\sigma=0.00577$. 
They are such that $\Delta=0.04$ and 0.02 respectively (see text). 
}

\end{figure}

The momentum distribution of a BEC, although rather sharp, has actually
a finite width, which induces a damping with time of the effects observed
in the ideal, zero-width case \cite{2008-dana-prl:QRRatchet-Exp}.
This is illustrated on Figs. \ref{fig:RQS-Emoy-t-balist} and
\ref{fig:RQS-Emoy-t-antir}, with numerical simulations made for a coherent initial momentum distribution, which is a Gaussian centered at zero ($\overline{P}(0)=0$), and whose root mean square is $\sigma$. This distribution is assumed to be so narrow ($\sigma\ll 1$), that it only populates the first Brillouin zone in a
 significant way,
\begin{eqnarray}
\left|\widetilde{\psi}_{\beta}(n)\right|^2 & = & \frac{1}{\sigma\sqrt{2\pi}}
\exp{\left(-\frac{\left(n+\beta\right)^2}{2\sigma^2}\right)} \nonumber\\
 & \approx & \frac{1}{\sigma\sqrt{2\pi}}
\exp{\left(-\frac{\beta^2}{2\sigma^2}\right)} \delta_{n0}\,.
\label{eq:RQS-thin-psi2-1}
\end{eqnarray}
Note that $\sigma\sim 10^{-2}$ for Bose-Einstein condensates.
Therefore, each quasimomentum $\beta$ is associated
to a plane wave centered at $\kbar\beta$, and the CSBW distribution
$\pi_{\xi,\beta}$ is thus independent of $\xi$,
\begin{eqnarray}
\pi_{\beta,\xi} = \frac{1}{\sigma\left(2\pi\right)^{3/2}}
\exp{\left(-\frac{\beta^2}{2\sigma^2}\right)} \,.
\end{eqnarray}

Figures \ref{fig:RQS-Emoy-t-balist} and \ref{fig:RQS-Emoy-t-antir} show that the behavior associated to the central quasimomentum is visible in the early dynamics. Then, it disappears after a characteristic time $\tau_{d}$, which decreases as the distribution width $\sigma$ increases.
Namely, Fig. \ref{fig:RQS-Emoy-t-balist} shows a numerical simulation of the
resonant case $\kbar=4\pi$ in which this modification manifests itself
as a transformation of the initial ballistic motion, characterized
by a quadratic increase of the average kinetic energy (see inset) into
a diffusive motion in later times $t>\tau_{d}$, characterized by
a linear increase of the average kinetic energy.
For the sharper initial distribution (in dashed lines), the change in the dynamics occurs slightly later, and the resulting diffusion is characterized by a higher rate.  
Fig. \ref{fig:RQS-Emoy-t-antir} shows the anti-resonant case $\kbar=2\pi$, whose characteristic oscillations of the average kinetic energy are damped.
The broader is the initial momentum distribution, the faster is the damping process.

Such changes between the early and the asymptotic dynamics has been observed for thermal gases, \textit{i.e.} incoherent mixtures of plane waves \cite{2007-saunders-pra:QR:FinTemp,2008-halkyard-pra:PowLaw}, whereas our simulations are made for Bose-Einstein condensates, \textit{i.e.} coherent superpositions of plane waves. For narrow momentum distributions ($\sigma\ll 1$), the dynamics is the same for coherent and incoherent superpositions. As the initial momentum and quasimomentum distributions are identical, the different momentum components evolve independently from each other, and so are not sensitive to any phase difference between them.
 
In order to give a simple picture of the dynamics, in what follows, we consider a square initial momentum distribution
\begin{equation}
\left|\widetilde{\psi}_{\beta}(n)\right|^2=\begin{cases}
\frac{1}{\Delta}\delta_{n0} & \textrm{for }|\beta|\le\frac{\Delta}{2}\\
0 & \textrm{otherwise.}\end{cases}
\end{equation}
The parameter $\Delta$ is such that the Gaussian and square momentum distributions have the same average kinetic energy, that is $\sigma=\Delta/\sqrt{12}$.
The corresponding CSBW distribution $\pi_{\xi,\beta}$ is
\begin{equation}
\pi_{\beta,\xi}=\begin{cases}
\frac{1}{2\pi\Delta} & \textrm{for }|\beta|\le\frac{\Delta}{2}\\
0 & \textrm{otherwise.}\end{cases}
\end{equation}
Applying Eq. (\ref{eq:RQS-E-op}) to this particular case, we can write the average kinetic energy in the form
\begin{eqnarray}
\overline{E}(t) & = & \overline{E}(0)+\frac{K^{2}}{4\Delta}
\int_{-\Delta/2}^{\Delta/2}d\beta f_{\beta}(t)\,,
\label{eq:RQS-E-thin-E-1}
\end{eqnarray}
which shows a filtering process in momentum space, associated to the function 
\begin{equation}
f_{\beta}(t)=\frac{\sin^{2}\left(\pi\ell\left(\beta+1/2\right)t\right)}{\sin^{2}\left(\pi\ell\left(\beta+1/2\right)\right)}\,,
\label{eq:RQS-thin-filter}
\end{equation}
characteristic of diffraction by a grating \cite{Wimberger:QuantRes:NL03}. As $f_{\beta}(t)$ becomes sharper with time, it better ``{}resolves'' the details of the 
initial momentum distribution, especially its finite width, which can explain the damping process observed on Figs. 
\ref{fig:RQS-Emoy-t-balist} and \ref{fig:RQS-Emoy-t-antir}.

\begin{figure}
\begin{centering}
\includegraphics[width=8cm]{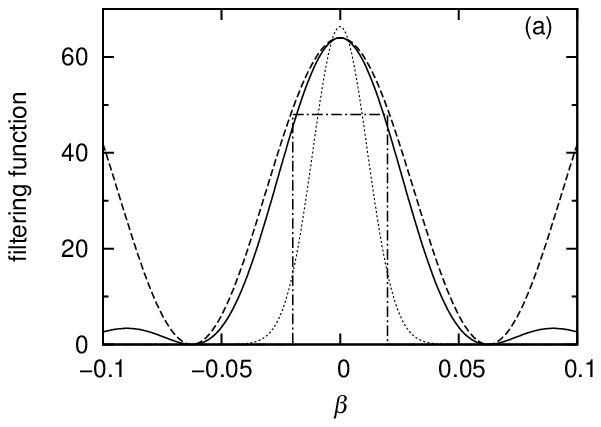}
\includegraphics[width=8cm]{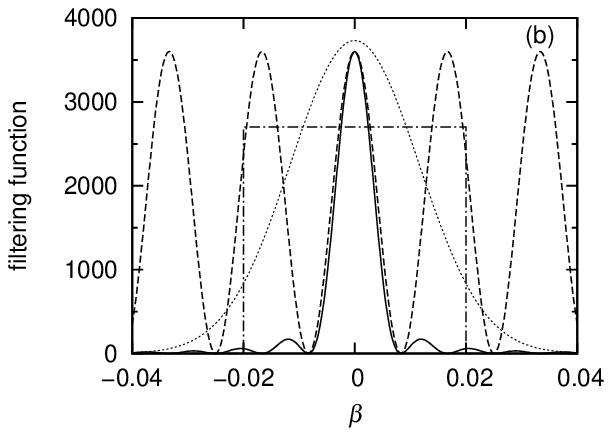}
\par\end{centering}

\caption{\label{fig:RQS-thin-filter}The filtering function $f_{\beta}(t)$
{[}see Eq. (\ref{eq:RQS-thin-filter})] in full lines, and its approximation
$f_{\beta}^{\prime}(t)$ {[}see Eq. (\ref{eq:RQS-thin-Approx-filter})]
in dashed lines: for (a) $t=8$ kicks, and (b) $t=60$ kicks. The
initial momentum distributions, of Gaussian shape ($\sigma=0.0115$) and of square shape ($\Delta=0.04$), are also plotted for comparison,
in dotted and dashed-dotted lines, respectively. They are multiplied by an arbitrary factor to be visible on the graph.}

\end{figure}

An alternative explanation of this damping effect can be obtained by considering
the behavior of a slice at position $\xi$ of the initial distribution.
After $t$ kicks, the parts of the distribution characterized by the
extremal values of $\beta=\pm\Delta/2$ have been translated, according
to (\ref{eq:SQR-rec-psi}), of $\pi\ell\left(\pm\Delta+1\right)t/2$.
The filtering process becomes significant when an initial slice $\xi$
has spanned one potential well (the {}``spatial'' first Brillouin
zone), which gives a damping time $\tau_{d}=2/\ell\Delta\sim\left(\ell\Delta\right)^{-1}\sim\left(\kbar\Delta\right)^{-1}$,
a trend confirmed by our numerical simulations.

Now let us turn to the specific case of Fig. \ref{fig:RQS-Emoy-t-balist}. 
Let us approximate the central lobe
$\left|\beta\right|t\le1$ of the filtering function $f_{\beta}(t)$
by
\begin{equation}
f_{\beta}'(t)=t^{2}\cos^{2}\left(\pi\beta t\right),\label{eq:RQS-thin-Approx-filter}
\end{equation}
which is shown in Figs. \ref{fig:RQS-thin-filter}(a) and (b)
for different times $t=8$ and $t=60$. If $t\le\Delta^{-1}$, the
central lobe completely encompasses the initial momentum distribution;
therefore the average kinetic energy is
\begin{eqnarray}
\overline{E}(t) & \approx & \overline{E}(0)+\frac{K^{2}t^{2}}{4\Delta}\int_{-\Delta/2}^{\Delta/2}d\beta\cos^{2}\left(\pi\beta t\right)\nonumber \\
 & \approx & \overline{E}(0)+\frac{K^{2}t^{2}}{8}\left(1+\frac{\sin\left(\pi\Delta t\right)}{\pi\Delta t}\right),\label{eq:RQS-thin-bal-E-1}
\end{eqnarray}
which shows, as expected, that the early dynamics is ballistic ($\overline{E}(t)\propto t^{2}$),
with a coefficient $K^{2}/4$, as for a plane wave. As $t$ approaches
$\Delta^{-1}$, Eq. (\ref{eq:RQS-thin-bal-E-1}) predicts that the
dynamics remains ballistic, but that the rate decreases from $K^{2}/4$
to $K^{2}/8$. This conclusion in not valid for $t>\Delta^{-1}$,
when the edges of $\pi_{\beta,\xi}$ fall outside of the central lobe
of the filtering function. We can then calculate the integral (\ref{eq:RQS-E-thin-E-1})
over the range of the central lobe, that is for $\left|\beta\right|\le1/2t$:
\begin{eqnarray}
\overline{E}(t) & \approx & \overline{E}(0)+\frac{K^{2}t^{2}}{4\Delta}\int_{-1/2t}^{1/2t}d\beta\cos^{2}\left(\pi\beta t\right),
\label{eq:RQS-thin-bal-E-2}
\end{eqnarray}
which yields a diffusive behavior ($\overline{E}(t)\propto t$)
\begin{equation}
\overline{E}(t)\approx\overline{E}(0)+\frac{K^{2}}{8\Delta}t,
\end{equation}
whose rate is inversely proportional to $\Delta$. This trend is also visible on
Fig. \ref{fig:RQS-Emoy-t-balist} for Gaussian momentum distributions.

In the anti-resonant case shown in Fig.~\ref{fig:RQS-Emoy-t-antir}, the energy,
although damped, oscillates for a time equal to a few $\Delta^{-1}$,
and finally ``freezes'' at $K^{2}/8$. To catch the most important
features of the dynamics, we will once again take an approximate form
for Eq.~(\ref{eq:RQS-E-thin-E-1}). Indeed, we can take $\sin^{2}\left\{ \pi\ell\left(\beta+1/2\right)\right\} \approx1$
over the width of the quasimomentum distribution $\Delta$. By expanding
$\sin^{2}\left\{ \pi\ell\left(\beta+1/2\right)t\right\} $, one gets
\begin{eqnarray}
\overline{E}(t) & \approx & \overline{E}(0)+\frac{K^{2}}{8}\left[1-\frac{1}{\Delta}\int_{-\Delta/2}^{\Delta/2}d\beta\cos\left\{ 2\pi t\left(\beta+1/2\right)\right\} \right]\nonumber \\
 & \approx & \overline{E}(0)+\frac{K^{2}}{8}\left[1-\left(-1\right)^{t}\frac{\sin\left(\pi\Delta t\right)}{\pi\Delta t}\right].
\label{eq:RQS-thin-E-antireso}
\end{eqnarray}
For $t\gg\Delta^{-1}$ the oscillations are completely damped and
the energy tends to its asymptotic value\begin{equation}
\overline{E}(t\to\infty)=\overline{E}(0)+\frac{K^{2}}{8}.\end{equation}

Incidentally, if one keeps in Eq. (\ref{eq:RQS-thin-E-antireso})
a general momentum distribution $\pi_{\beta,\xi}$, the equation shows
that the kinetic energy is proportional to the cosine-Fourier series
of the momentum distribution (see Fig. \ref{fig:RQS-Emoy-t-antir}). 
Using the properties of Fourier series,
we can {}``invert'' Eq. (\ref{eq:RQS-thin-E-antireso}) to obtain
\begin{equation}
\left|\widetilde{\psi}_0(\beta)\right|^2 \approx \sum_{s=1}^{t}\frac{4}{K^{2}}\left(\overline{E}(0)-\overline{E}(t)\right)\cos\left(2\pi t\left(\beta+\frac{1}{2}\right)\right),\label{eq:SQR-InitMomDistr}
\end{equation}
which yields an approximate expression for the initial distribution.
 Since it appears as a Fourier series, the accuracy of this approximation
increases when $t\to\infty$. This result is valid for narrow momentum
distributions that contain no ballistic quasimomentum. As ballistic
quasimomenta are more numerous with increasing $\kbar$, the phenomenon
is more likely to be observed for $\kbar=2\pi$.

\section{Broad initial momentum distribution}\label{sub:Broad}

For the sake of completeness, we consider now the trivial case of
an initial momentum distribution that is a few times larger than the
Brillouin zone. This case leads to a well-known diffusive behavior,
\begin{equation}
\overline{E}(t)=\overline{E}(0)+\frac{K^{2}}{4}t\,,
\end{equation}
which was demonstrated with thermal gases \cite{Wimberger:QuantRes:NL03,2007-saunders-pra:QR:FinTemp,2008-halkyard-pra:PowLaw}. In this section, we use our approach to demonstrate that the diffusion can be observed both with coherent and incoherent superpositions of plane waves.

The momentum-space wavepacket associated with a coherent, say Gaussian, superposition of plane waves is
\begin{equation}
\widetilde{\psi}_{\beta}(n) = \frac{e^{-i(n+\beta)\phi}}{\sqrt{\sigma\sqrt{2\pi}}}
\exp{\left(-\frac{\left(n+\beta\right)^2}{4\sigma^2}\right)} \,,
\end{equation} 
where $\sigma\gg 1$.  The corresponding CSBW distribution is approximately uniform is $\beta$, but strongly localized about $\xi=\phi$. Eq. (\ref{eq:RQS-map-X}) shows that, after one kick, the part of the CSBW distribution corresponding to quasimomentum $\beta$ moves to $\phi+2\pi\ell\left(\beta+1/2\right)$, which means that the CSBW distribution covers a whole potential well. Therefore, the average momentum does not change, and the average kinetic energy is increased by $\int d\beta$ $K^{2}\sin^{2}(\xi+2\pi\ell\beta)/2=K^{2}/4$, hence the diffusive behavior.

For incoherent mixtures, there is no well-defined phase $\phi$ between the different momentum components. By applying Eq.~(\ref{eq:pi-rho}) and by supposing that all quasimomenta are equally populated, we can write $\pi_{\beta,\xi}\approx\left(2\pi\right)^{-1}$. The reasoning made for a coherent superposition of plane waves can be applied for a given slice $\xi$, hence the diffusive behavior characterized by the rate $K^2/4$, which is then not modified by the average over $\xi$.

\section{Conclusion}\label{sec:Conclusion}

In this paper, we have applied the analysis of quantum resonances
in position space given in \cite{AP:QuantumRes:PRA08} to cases that
are met in the everyday laboratory life. Our interpretation of the
simple quantum resonances, based on the local momentum transferred
from the kicks, can also be applied to wave functions that are not
sharply localized in the {}``spatial first Brillouin zone''. As shown
above, this allows one to build intuitive and easily understandable
images of the phenomena that are often simpler than the corresponding
explanation in the momentum representation.
We have applied our approach to both coherent and incoherent superpositions of plane waves, which stand for Bose-Einstein condensates and thermal gases, respectively. We have demonstrated that the two types of superpositions induce the same dynamical behavior, in the two limiting cases of narrow and broad initial momentum distributions. By contrast, we expect significant differences to appear if the width of the distribution is comparable to the first Brillouin zone.

The case of high-order quantum resonances has not been addressed here.
For $\kbar=\pi$, the initial wave packet reconstructs into two sub-packets
separated by half the lattice step. When a kick is applied, the momentum
shift is due to the local potential gradient at the place of the sub-packets
\cite{AP:QuantumRes:PRA08}. However, the weights of the sub-packets
depend on their phase difference, giving birth to a complex interfering
pattern after a few kicks. The evolution of the average kinetic energy
reflects this {}``history'', and therefore cannot be explained in
terms of the simple arguments developed in this paper.

An interesting prospect of the present work is its application to
the case where interactions between atoms are present (e.g. one uses
a dense Bose-Einstein condensate). A first experimental study of such
as system has been made by Raizen and co-workers \cite{Raizen:QRBEC:PRL06},
and it was demonstrated in \cite{Monteiro:QRBEC:PRL09} that for a
weak interaction strength, the resonant value of $\kbar$ is shifted.
The latter work also shows that the diffusion coefficient presents
sharp edges as the non linearity is varied. A clear picture of the
physics underlying these observations is still missing. A kind of
nonlinear \emph{optical} Talbot effect has been observed recently
in an optical system \cite{Xiao:NLOptTalbotEffect:PRL10}, which hints
that our approach could be generalized to this much more complex case.

\bibliographystyle{unsrt}
\bibliography{ArtDataBase}

\end{document}